\documentclass[manuscript]{acmart}

\pdfoutput=1
\setcopyright{none}
\usepackage{multirow}
\usepackage[normalem]{ulem}
\usepackage{xcolor}
\useunder{\uline}{\ul}{}
\usepackage{xcolor,colortbl}
\usepackage{booktabs}
\usepackage{float}
\usepackage{caption}

\usepackage{titlesec}

\titlespacing\section{0pt}{12pt plus 4pt minus 2pt}{0pt plus 2pt minus 2pt}
\titlespacing\subsection{0pt}{12pt plus 4pt minus 2pt}{0pt plus 2pt minus 2pt}
\titlespacing\subsubsection{0pt}{12pt plus 4pt minus 2pt}{0pt plus 2pt minus 2pt}
\titlespacing\paragraph{0pt}{12pt}{12pt}

\usepackage[T1]{fontenc}
\usepackage[font=small,labelfont=bf,tableposition=top]{caption}

\usepackage[font=small,labelfont=bf]{caption}

\AtBeginDocument{%
  \providecommand\BibTeX{{%
    \normalfont B\kern-0.5em{\scshape i\kern-0.25em b}\kern-0.8em\TeX}}}

\copyrightyear{2022} 
\acmYear{2022} 
\setcopyright{rightsretained} 
\acmConference[UbiComp/ISWC '22 Adjunct]{Proceedings of the 2022 ACM International Joint Conference on Pervasive and Ubiquitous Computing}{September 11--15, 2022}{Cambridge, United Kingdom}
\acmBooktitle{Proceedings of the 2022 ACM International Joint Conference on Pervasive and Ubiquitous Computing (UbiComp/ISWC '22 Adjunct), September 11--15, 2022, Cambridge, United Kingdom}
\acmDOI{10.1145/3544793.3560333}
\acmISBN{978-1-4503-9423-9/22/09}

\begin{document}

\title{Privacy-Patterns for IoT Application Developers}


 \author{Nada Alhirabi}
     \affiliation{%
 \institution{Cardiff University}
  \city{Cardiff}
  \country{UK}
}
 \affiliation{%
 \institution{King Saud University}
  \city{Riyadh}
  \country{Saudi Arabia}
}
 \email{alhirabin@cardiff.ac.uk}

 \author{Stephanie Beaumont}

 \affiliation{%
  \institution{My Data Fix Ltd}
  \city{London}
\country{UK}
}
 \email{stephaniebeaumont@mydatafix.com}

 \author{Omer Rana}
    \affiliation{%
 \institution{Cardiff University}
  \city{Cardiff}
  \country{UK}
}
 \email{ ranaof@cardiff.ac.uk}
 
  \author{Charith Perera}
    \affiliation{%
 \institution{Cardiff University}
  \city{Cardiff}
  \country{UK}
}
 \email{pererac@cardiff.ac.uk}
 
 

\renewcommand{\shortauthors}{Alhirabi et al.}

\begin{abstract}
Designing Internet of things (IoT) applications (apps) is challenging due to the heterogeneous nature of the systems on which these apps are deployed. Personal data, often classified as sensitive,  
may be collected and analysed by IoT apps, where data privacy laws are expected to protect such information. 
Various approaches already exist to support privacy-by-design (PbD) schemes, enabling developers to take data privacy into account at the design phase of application development. However, developers are not widely adopting these approaches because of understandability and interpretation challenges. A limited number of tools currently exist to assist developers in this context -- leading to our proposal for ``PARROT" (PrivAcy by design tool foR inteRnet Of Things). PARROT supports a number of techniques to enable PbD techniques to be more widely used. 
We present the findings of a controlled study 
and discuss how this privacy-preserving tool increases the ability of IoT developers to apply privacy laws (such as GDPR) and privacy patterns. 
Our students demonstrate that the PARROT prototype tool increases the awareness of privacy requirements in design and increases the likelihood of the subsequent design to be more cognisant of data privacy requirements.

\end{abstract}



\maketitle

\section{Introduction}

Internet of Things (IoT) applications generate and process a large amount of data that are transmitted between devices. 
As the size and frequency of this data increase, an efficient architecture is needed to manage and process this data~\cite{Kumar2019}. Many efforts have been made to support privacy in the early stage of software development, such as Privacy-by-Design (PbD) principles by Cavoukian \cite{cavoukian2009privacy}.
However, many developers are unaware of the potentially  significant privacy issues in an online context -- finding it time-consuming and challenging to understand privacy  policies and their implications for their work \cite{cranor2006user}. 
Moreover, privacy concerns for a specific app design or implementation are rarely discussed by developers \cite{Li2021}.
This indicates a need for a privacy tool to reduce the operational and implementation gap between software developers and privacy requirements \cite{Alhirabi2021}. 

The PARROT tool offers intuitive and user-friendly interfaces to assist and educate software developers on how to learn and include privacy in their system design \cite{AlhirabiDemo2022}. Initially, the tool was built for the highly regulated domain of healthcare. Since then, we have added more use cases, such as smart homes and multi-cloud systems, to test different sensors and privacy challenges such as managing advertisements, cookies and payments.



\section{Architecture and Implementation}
PARROT is an interactive prototype tool that was implemented using Sirius (eclipse.org/sirius), a domain-specific modelling tool, to test the effectiveness of privacy by design principles. %
We have assessed the gaps and challenges that developers usually face when planning to consider privacy by design. Therefore, this prototype is intended to act as a privacy assistant for software developers. To improve visual support, we used a simple visual notation based on: Size, Shape, and Colour \cite{Moody2010}. We co-designed this software tool collaboratively with a privacy lawyer and other privacy professionals to take their differing perspectives into account. We constructed the tool based on the six Cavoukian PbD principles \cite{cavoukian2009privacy}, which are: \textit{(1) Privacy requirements intrinsic in design and analysis,  (2) Privacy embedded in the design, (3) Full functionality, (4) End-to-end security, (5) Visibility and transparency,} and \textit{(6) Respect for user privacy}.

\section{Evaluation}
\label{sec:Methodology}

\begin{figure}[h]
	\includegraphics[scale=.37]{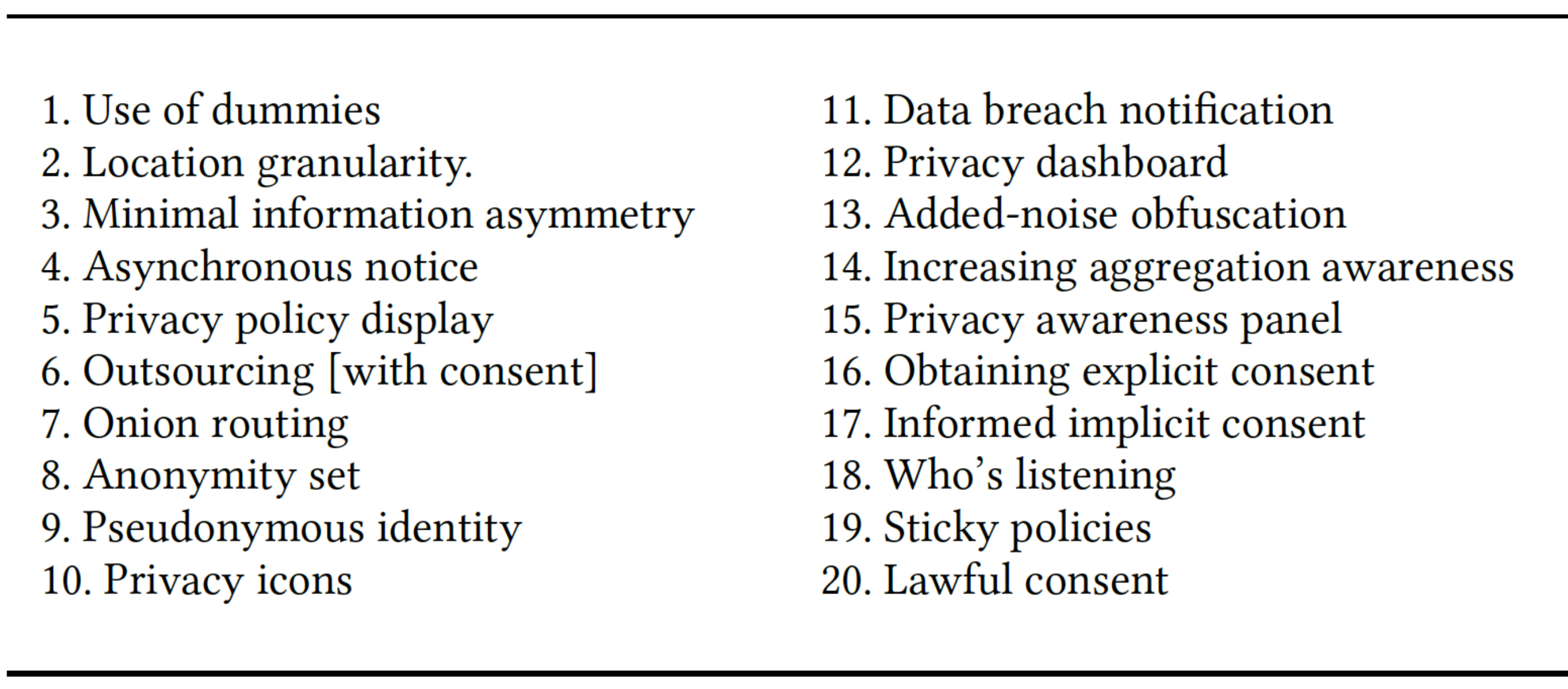}
      \caption{List of the aplicable privacy patterns from sources:   (privacypatterns.org) and  (privacypatterns.eu)}
	\label{table:privacy patterns}
\end{figure}

We conducted a controlled lab study to answer the following research questions: (RQ1) does the tool enable the design of privacy-aware IoT applications for less regulated domains, in comparison to a highly regulated domain such as healthcare? (RQ2) does the tool help increase awareness in software developers about privacy-preserving measures such as privacy patterns. 
Since software design is typically a collaborative activity, participants worked in pairs.\\
\textbf{\textit{Recruitment:}}
We recruited participants through the University email group targeting computer science  students (UG, PG taught and PG Research)  who worked on IoT applications for at least a year \cite{host2000using}. 
We hired 12 participants for the study where each participant was given a voucher after completing the study.\\
\textbf{\textit{Evaluation sessions:}}
All the study sessions were conducted online, where each study session lasted between 1.5-2 hours.
We performed between-subjects evaluation to test if PARROT developers are able to create more privacy-preserving IoT designs. We also tested the potential increase in privacy awareness of participants.
In the study, each participant was allocated to one of the two conditions (using or not using PARROT). Twelve participants were divided into an experimental(E) and a control(C) group. Both groups had 6 participants each; both groups involved participants working in pairs. 
At the beginning, both groups were given a 20-minutes introduction to privacy, followed by a tutorial on Mural for Group C only and PARROT for Group E only. The participants were then given a list of 20 privacy patterns that were picked based on their applicability to the use case (Figure \ref{table:privacy patterns}).
We asked the control group (C) to use the Mural tool to do the design task for the smart home scenario considering privacy rules and privacy patterns.  
 The experimental group (E) performed the same task but using the PARROT tool. Both groups had an exit questionnaire for ten minutes at the end of the session.\\
\textbf{\textit{Data scoring:}}
To evaluate the overall privacy principle score, we assigned a score 
 for each principle by the lawyer as 3: if privacy is considered, the issue is identified and the solution is correct; 2: if privacy is considered, and the issue is identified; 1: if privacy is considered; and 0: if no privacy requirement is considered. We also assigned a score for each privacy pattern as 0: if no pattern was considered; 1: if privacy pattern was considered overall but not in a reasonable place;
2: if a privacy pattern was considered in a reasonable place. A privacy patterns expert was consulted to reduce researcher basis. Then, we sum the score up for all the patterns and principles to have the total score for each one.


\section{Quantitative and Qualitative Results}
To evaluate the study results, Kruskal-Wallis test was performed to determine if there was a statistically significant difference between the groups (E and C). Both privacy principles (p-value = $0.0463<0.05$) and privacy patterns (p-value  

\begin{figure}[h]
	\includegraphics[scale=.28]{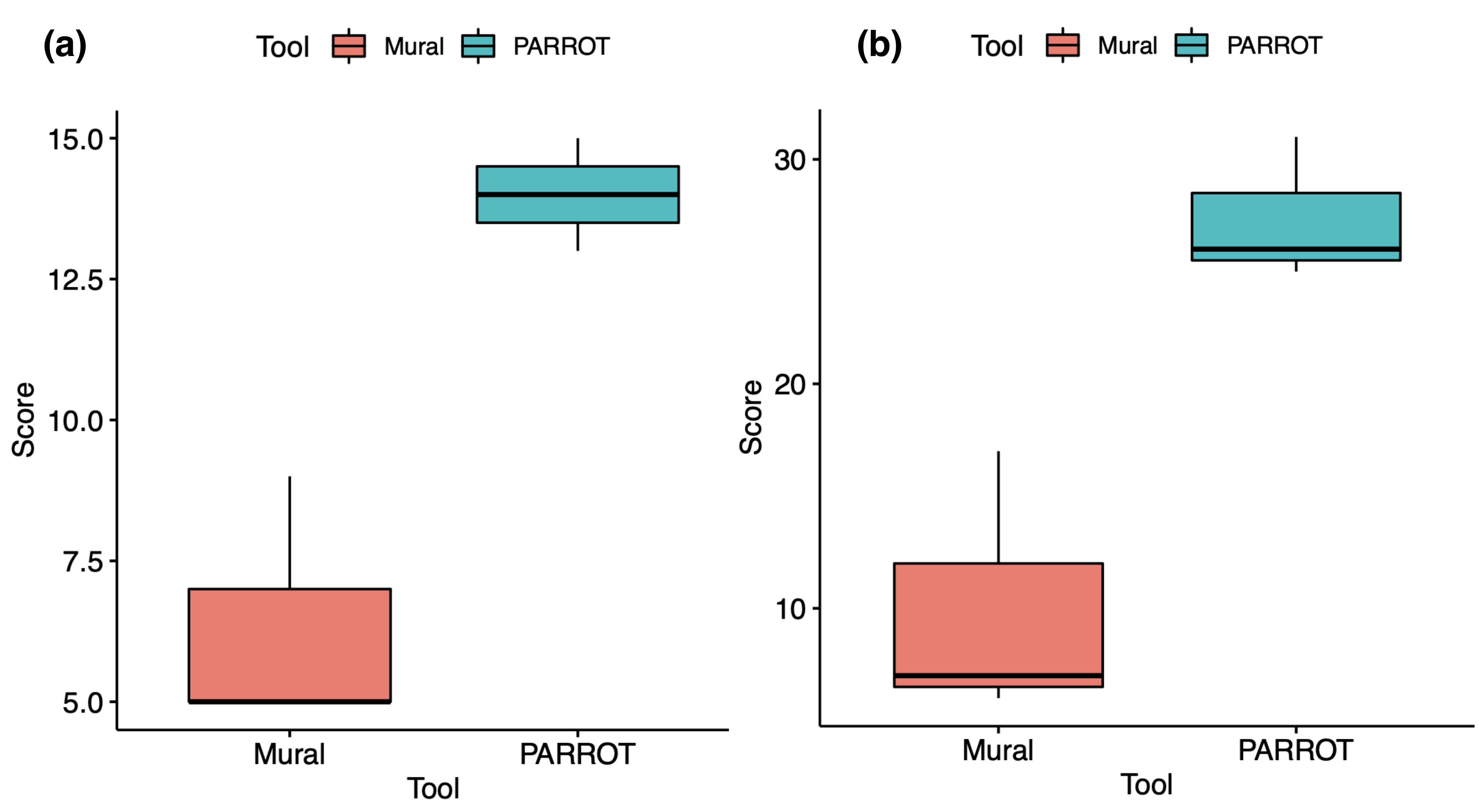}
	   \caption{Mean rates of privacy principles scores in Mural and PARROT. (b) Mean rates of privacy patterns in Mural and PARROT.}
	\label{Fig:Study1ScorsPlotBox}	
\end{figure}

  \noindent
=$0.04953 < 0.05$) revealed a significant difference. For posthoc test, Dunn Test was used to test if there is statistical difference  between Mural and PARROT. We observed a significant difference
for both  privacy principles (p-value=0.046 30159) and privacy patterns (p-value =0.04953461), as shown in Figure~\ref{Fig:Study1ScorsPlotBox}.
We also performed a qualitative analysis to have more insight into participants' thoughts and ideas. 
We discussed how the tool helps integrate privacy principles and patterns with the lawyer and the participants.
The privacy lawyer said that PARROT was able to include privacy-specific design components into the IoT application \textit{``from the beginning rather than retrospectively"} from a privacy compliance perspective. In addition, several participants expressed their preference for the visual representation of PARROT. 
 For example, Pair 2 said, \textit{``the generated colours are helpful to flag any privacy issue immediately... I think it helps to rethink the question again"}. 
 Pairs 1, 4 and 5 believed PARROT could help people who do not have any privacy background to understand it in a short period. Pair 4 said \textit{``I definitely struggle to understand and apply privacy and privacy patterns because there are many different documents, laws and IoT devices... PARROT will tell you already what privacy needs to be fulfilled for that node which is super useful, in my opinion...you don't have to start researching about it"}. \\
 Pair 1, 4, 5 and 6 said that the questions led them to think about things they had not considered previously. For example, Pair 1 said, \textit{"the questions and visual presentation make me aware of little things... presenting privacy when you are setting up is very helpful."}. Pair 4  stated that \textit{``the variety of questions you got asked makes you think of how you can make this correctly"}. Pair 5 said, \textit{``the questions help me to think more about the data subject perspective, not the problem owner only"}.

\section{Conclusion and Future Plan}
This paper presents and discusses the findings of  PARROT, an interactive prototype tool to assist developers with privacy. Our participants demonstrated how the PARROT prototype tool helps to embed privacy principles and increases their awareness of privacy patterns. We plan to add more use cases and features, such as showing the overall privacy score of the design and adding menus that include all the applicable privacy patterns in each part of the design.


\bibliographystyle{plain}
\bibliography{PARROTPoster}
\end{document}